\documentclass{webofc}
\usepackage[varg]{txfonts}   
\usepackage{tikz}

\usepackage{amsmath}
\usepackage{amsfonts}
\usepackage{amssymb}
\usepackage{physics}
\usepackage{slashed}
\usepackage{simplewick}
\usepackage{mathtools}

\usepackage{subcaption}
\usepackage{caption}

\usepackage{color}
\usepackage{siunitx}

\usepackage{feynmp-auto}

\usepackage{graphicx}
\usepackage{float}

\begin{document}

\begin{fmffile}{feynDiags}
\title{Studying the production mechanisms of light meson resonances in two-pion photoproduction}
%
%

\author{
\firstname{Łukasz} \lastname{Bibrzycki}\inst{1}
\and
\firstname{Nadine} \lastname{Hammoud}\inst{2}
\and
\firstname{Vincent} \lastname{Mathieu}\inst{3,4}
\and
\firstname{Robert J.} \lastname{Perry}\inst{3}\thanks{\email{perryrobertjames@gmail.com}} 
\and
\firstname{Adam P.} \lastname{Szczepaniak}\inst{5,6,7}
~ for the JPAC Collaboration
}

\institute{Faculty of Physics and Applied Computer Science, AGH University of Krakow,
al. A. Mickiewicza 30, 30-059 Kraków, Poland
\and
Institute of Nuclear Physics, Polish Academy of Sciences, PL-31-342 Krak\'ow, Poland
\and
Departament de F\'isica Qu\`antica i Astrof\'isica and Institut de Ci\`encies del Cosmos, Universitat de Barcelona,  E-08028 Barcelona, Spain
\and
Departamento de F\'isica Te\'orica, Universidad Complutense de Madrid and IPARCOS, E-28040 Madrid, Spain
\and
Theory Center, Thomas Jefferson National Accelerator Facility, Newport News, VA 23606, USA
\and
Center for Exploration of Energy and Matter, Indiana University, Bloomington, IN 47403, USA
\and
Department of Physics, Indiana University, Bloomington, IN 47405, USA
}

\abstract{%
A theoretical model of two-pion photoproduction is presented.
The model encodes the prominent $\rho(770)$ resonance and the expected leading background contribution coming from the Deck mechanism. To validate the model, angular moments are computed and compared with the CLAS dataset. After fitting a number of free parameters, the model provides a good description of the data.
}
\maketitle

\section{Introduction}
\label{intro}
Hadron photoproduction is an essential class of experimental measurements which give important information on the spectroscopic and structural nature of hadrons in quantum chromodynamics (QCD). Given the clear value in studying hadronic photoproduction, a range of dedicated experiments continue to investigate these processes. Two-pion photoproduction is a well-studied example of hadron photoproduction. The most recent public data for this process come from the CLAS experiment~\cite{CLAS:2009ngd}, and it is anticipated that further data will be published by both CLAS12 and GlueX. Thus it will be important to have accurate hadronic models to compare with these new, high-precision experimental results. 

This proceedings focuses on a theoretical description of two (charged) pion photoproduction at low invariant mass of the $\pi\pi$ system. In this kinematic region, the data exhibits a clear peak which may attributed to the presence of the $\rho(770)$ resonance. In addition to this prominent resonance, several $S$-wave resonances are expected to contribute to the cross section in this kinematical region. In particular, the model incorporates resonant contributions from the $\sigma/f_0(500)$, and the $f_0(980)$. The proceedings is organized as follows. The kinematics for the process are introduced in Sec.~\ref{sec:kinematics}, the components of the model are described in Sec.~\ref{sec:model} and the predictions for the angular moments are presented in Sec.~\ref{sec:results}.

\section{Kinematics and Preliminaries}
\label{sec:kinematics}
In this proceedings, a theoretical model for the process
\begin{equation}
\vec{\gamma}(q,\lambda_\gamma)+p(p_1,\lambda_1)\to \pi^+(k_1)+\pi^-(k_2)+p(p_2,\lambda_2)\,,
\end{equation}
is described. Observables are computed in the the $\pi^+\pi^-$ helicity rest frame. In this frame, the $\pi^+\pi^-$ system is at rest ($\mathbf{k}_1=-\mathbf{k}_2$) and the recoiling proton ($\mathbf{p}_2$) defines the negative $z$-axis. The reaction plane is defined by the photon, the target and the recoiling proton. One can use this plane to define the $x$-$y$ plane. Then the unit vector normal to this defines the $y$-axis. With this choice, $\Omega^\text{H}=(\theta^\text{H},\phi^\text{H})$ defines the angles of the $\pi^+$ with respect to the $z$-axis. 

The scalar amplitudes in such a $2\to3$ process are maximally described by five independent kinematic variables. Thus in addition to the two angles for the $\pi^+$, the following kinematic invariants are commonly used:
\begin{align}
s&=(p_1+q)^2=(p_2+k_1+k_2)^2\,,
\\
t&=(p_1-p_2)^2=(k_1+k_2-q)^2\,,
\\
s_{12}&=(k_1+k_2)^2=(p_1-p_2+q)^2=m_{12}^2\,,
\\
u_i&=(q-k_i)^2\,;~~~~i=1,2.
\end{align}
This set of kinematic variables, $(s,t,s_{12},\Omega^\text{H})$ is complete, in the sense that all other kinematic invariants may be computed by knowing these five.
The notation for the phase-space and intensity is taken from Ref.~\cite{Mathieu:2019fts}. The observables of interest are the \textit{angular moments}, which are defined as integral moments of the cross section:
\begin{equation}
\expval{Y_{LM}}(s,t,m_{12})=\sqrt{4\pi}\int d\Omega\, \frac{d\sigma}{dtdm_{12}d\Omega^\text{H}}\text{Re}\,Y_{LM}(\Omega^\text{H})\,.
\end{equation}

\section{Description of the Model}
\label{sec:model}
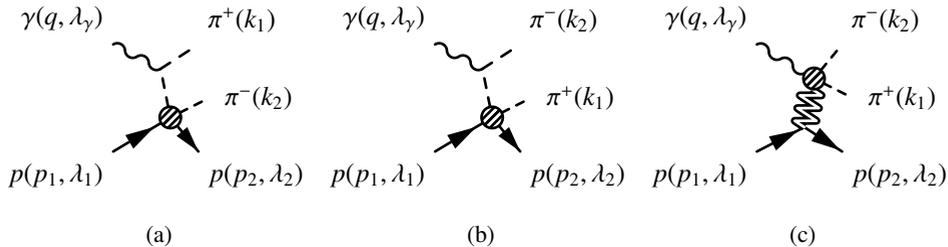
\begin{figure}
\vspace{20pt}
\begin{subfigure}{0.32\textwidth}
\centering
\begin{fmfgraph*}(40,40)
\fmfleft{l1,l2}
\fmfright{r1,r2,r3}
\fmf{fermion}{l1,v1,r1}
\fmf{photon}{l2,v2}
\fmf{dashes}{v1,r2}
\fmf{dashes}{v1,v2,r3}
\fmfblob{0.2w}{v1}
\fmflabel{$p(p_1,\lambda_1)$}{l1}
\fmflabel{$\gamma(q,\lambda_\gamma)$}{l2}
\fmflabel{$p(p_2,\lambda_2)$}{r1}
\fmflabel{$\pi^-(k_2)$}{r2}
\fmflabel{$\pi^+(k_1)$}{r3}
\end{fmfgraph*}
\vspace{20pt}
\caption{\label{fig:deck1}}
\end{subfigure}
\begin{subfigure}{0.32\textwidth}
\centering
\begin{fmfgraph*}(40,40)
\fmfleft{l1,l2}
\fmfright{r1,r2,r3}
\fmf{fermion}{l1,v1,r1}
\fmf{photon}{l2,v2}
\fmf{dashes}{v1,r2}
\fmf{dashes}{v1,v2,r3}
\fmfblob{0.2w}{v1}
\fmflabel{$p(p_1,\lambda_1)$}{l1}
\fmflabel{$\gamma(q,\lambda_\gamma)$}{l2}
\fmflabel{$p(p_2,\lambda_2)$}{r1}
\fmflabel{$\pi^-(k_2)$}{r3}
\fmflabel{$\pi^+(k_1)$}{r2}
\end{fmfgraph*}
\vspace{20pt}
\caption{\label{fig:deck2}}
\end{subfigure}
\begin{subfigure}{0.32\textwidth}
\centering
\begin{fmfgraph*}(40,40)
\fmfleft{l1,l2}
\fmfright{r1,r2,r3}
\fmf{fermion}{l1,v1,r1}
\fmf{photon}{l2,v2}
\fmf{dashes}{v2,r2}
\fmf{dbl_zigzag}{v1,v2}
\fmf{dashes}{v2,r3}
\fmfblob{0.2w}{v2}
\fmflabel{$p(p_1,\lambda_1)$}{l1}
\fmflabel{$\gamma(q,\lambda_\gamma)$}{l2}
\fmflabel{$p(p_2,\lambda_2)$}{r1}
\fmflabel{$\pi^-(k_2)$}{r3}
\fmflabel{$\pi^+(k_1)$}{r2}
\end{fmfgraph*}
\vspace{20pt}
\caption{\label{fig:res}}
\end{subfigure}
\vspace{-10pt}
\caption{Dominant contributions to two-pion photoproduction at small momentum transfer. \ref{fig:deck1} and \ref{fig:deck2} represent the Deck topology and \ref{fig:res} represents the resonant contribution.
With this approximation, it is possible to relate this $2\to3$ process to $2\to2$ processes.
}\label{fig:model}
\end{figure}

The coarse features of the angular moments, $\expval{Y_{LM}}$, measured in two-pion photoproduction can be understood as the result of two competing mechanisms, which will be referred to as the resonant and continuum pieces, respectively. In this model, a clear resonant peak can be observed in the $m_{12}$ dependence which may be attributed to the $\rho(770)$ resonance. In addition, the PDG lists several resonances below 1~\si{GeV} which are expected to contribute to this process. Each of these resonances is incorporated by postulating a Regge production amplitude. 
The leading continuum contribution is attributed to diffractive scattering of the photon on the target nucleon. As a result of the long-range character of the interaction, one anticipates the dominance of one-pion exchange. By factorizing the one-pion-exchange (see Fig.~\ref{fig:model}), one sees that in the resulting $2\to2$ subprocess may be related to elastic $\pi N$ scattering. In the following subsections, further details are given on both pieces of the model.

\subsection{Resonant Model}
\label{sec:model:resonant}
It is assumed that resonances in the upper vertex may be expressed as the product of a production and decay amplitude, with the line-shape described by a Breit-Wigner distribution:
\begin{equation}
\mathcal{M}_{\lambda_1\lambda_2\lambda_\gamma}(s,t,s_{12},\Omega^\text{H})=S_\text{BW}(s_{12})\sum_{\lambda_R}\mathcal{M}_{\lambda_1\lambda_2\lambda_\gamma\lambda_R}^{\gamma N\to R N}\Gamma_{\lambda_R}^{R\to\pi\pi}\,,
\end{equation}
where the energy-dependent Breit-Wigner distribution takes the form suggested in Ref.~\cite{ParticleDataGroup:2022pth}. The model incorporates the known light meson resonances up to 1~\si{GeV} relevant for this channel. In particular, the model contains the $f_0(500)/\sigma$, $\rho(770)$ and $f_0(980)$. 
The photon energies for the CLAS dataset are of the order $E_\gamma\sim 3~\si{GeV}$. This corresponds to an $s$ of $s\sim 7~\si{GeV^2}$, while the range of momentum transfer values are as large as $t\sim -1~\si{GeV^2}$. In this kinematic region, the quasi $2\to2$ process may be written in terms of a sum of $t$-channel Reggeon exchanges, $E$. Regge theory predicts that the amplitude factorizes into the product of a Regge propagator, $R^E(s,t)$, and two vertices, $T^E$ and $B^E$ which are functions of the momentum transfer alone. Thus
\begin{equation}
\mathcal{M}_{\lambda_1\lambda_2\lambda_\gamma\lambda_R}^{\gamma N\to R N}(s,t,s_{12})=\sum_{E}T_{\lambda_\gamma\lambda_R}^E(t;s_{12})R^E(s,t)B_{\lambda_1\lambda_2}^E(t)\,.
\end{equation}
The form of the Regge propagator depends on the naturalness of the exchanged Reggeon:
\begin{equation}
R^E(s,t)=
\begin{dcases}
\frac{1+e^{-i\pi \alpha^E(t)}}{\sin\pi\alpha^E(t)}\bigg(\frac{s}{s_0}\bigg)^{\alpha^E(t)},~~ &E=\pi,\eta,\text{ (unnatural)}\,,
\\
\frac{\alpha^E(t)}{\alpha^E(0)}\frac{1+e^{-i\pi \alpha^E(t)}}{\sin\pi\alpha^E(t)}\bigg(\frac{s}{s_0}\bigg)^{\alpha^E(t)},~~ &E=\mathbb{P},f_2,a_2\text{ (natural)}\,,
\end{dcases}
\end{equation}
where $\alpha^E(t)=\alpha_0^E+\alpha_1^E t$ is the Regge trajectory, and  $s_0=1~\si{GeV^2}$ is a mass-scale introduced to fix the dimensions. The $t$-dependence of the vertices is not predicted from Regge theory, and must be inferred from data, or taken from a model. In this case, the above Regge model is matched to a one-particle exchange model. This matching fixes the form of the $T_{\lambda_\gamma\lambda_R}(t;s_{12})$, $B_{\lambda_1\lambda_2}(t)$ and $\Gamma_{\lambda_R}^{R\to\pi\pi}$ vertices. The $P$-wave amplitude may be found in Ref.~\cite{Lesniak:2003gf}. Flexibility is ensured by partial wave projecting each resonant amplitude, and then reweighting each partial wave with a free parameter $g_{LM}$, which in general is complex. These parameters are fit to the experimental data. 

\subsection{Background Model}
\label{sec:model:continuum}
The Deck Mechanism, which forms the continuum contribution to the model may be understood as a two-step process. One imagines the incoming photon first decaying into two (off-shell) pions. Then, one of these pions recoils elastically against the nucleon target, producing the desired $p\pi^+\pi^-$ final state. The amplitude for the gauge invariant Deck Model may be written as~\cite{Bibrzycki:2018pgu,Bibrzycki:2019stc,Bibrzycki:2020zts,Bibrzycki:2020gdp,Hammoud:2022new}
\begin{equation}
\begin{split}
\mathcal{M}_{\lambda_1\lambda_2\lambda_\gamma}^\text{GI Deck}(s,t,s_{12},\Omega)&=\sqrt{4\pi\alpha}\bigg[\bigg(\frac{\epsilon(q,\lambda_\gamma)\cdot k_1}{q\cdot k_1}-\frac{\epsilon(q,\lambda_\gamma)\cdot(p_1+p_2)}{q\cdot (p_1+p_2)}\bigg)\beta(u_1){M}_{\lambda_1\lambda_2}^-(s_2,t;u_1)
\\
&-\bigg(\frac{\epsilon(q,\lambda_\gamma)\cdot k_2}{q\cdot k_2}-\frac{\epsilon(q,\lambda_\gamma)\cdot(p_1+p_2)}{q\cdot(p_1+p_2)}\bigg)\beta(u_2){M}_{\lambda_1\lambda_2}^+(s_1,t;u_2) \bigg]\,,
\end{split}
\end{equation}
where $\beta(u_i)=\exp((u_i-u_i^\text{min})/\Lambda_\pi^2)$ are hadronic form factors introduced to suppress the Born term pion propagator for the one-pion exchange at large $u_i$, $\Lambda_\pi=0.9~\si{GeV}$ and ${M}_{\lambda_1\lambda_2}^\pm$ is the scattering amplitude for the process $p+\pi^{*\pm }\to p+\pi^\pm$. Gauge invariance is ensured by the introduction of a phenomenological contact term $V^\mu$~\cite{Pumplin:1970kp}. Note that although this is a binary amplitude, the virtuality of the initial pion, $u_i$ implies that this amplitude is dependent on three kinematic invariants. In the limit that $u_i\to m_\pi^2$, these amplitudes may be related to elastic $\pi^\pm p$ scattering, for which there is a wealth of experimental and theoretical information. In this work, the $\pi N$ scattering amplitudes of Ref.~\cite{Mathieu:2015gxa} are used. They incorporate the SAID partial wave parameterization~\cite{Workman:2012hx} at low energies, but interpolate to a Regge description at large center of mass energies.

\begin{figure}[t]
\centering
\includegraphics[scale=0.25]{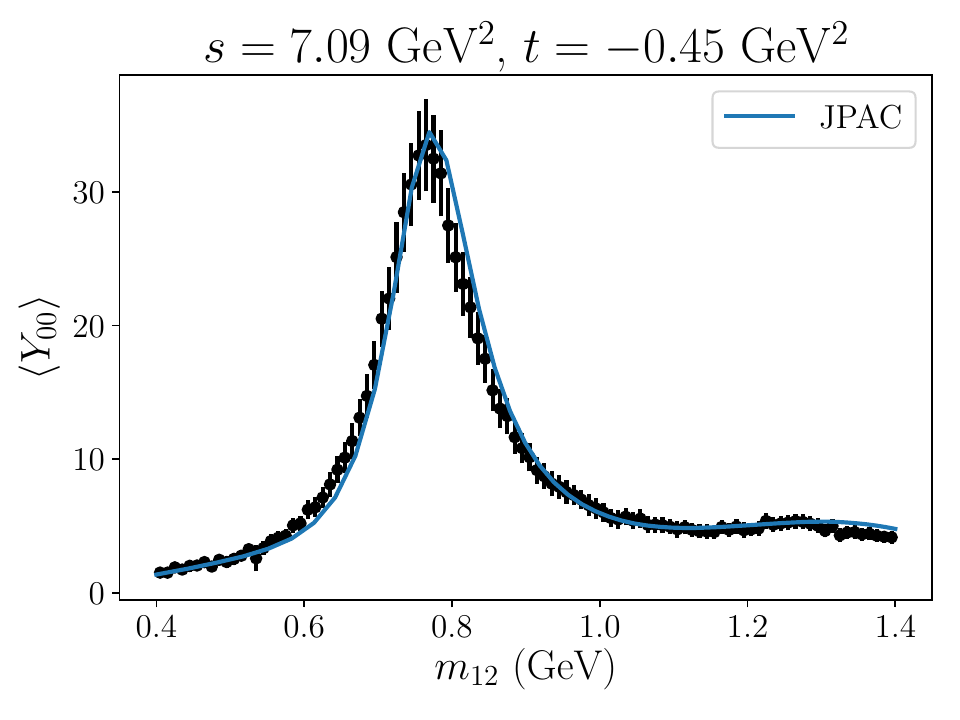}
\includegraphics[scale=0.25]{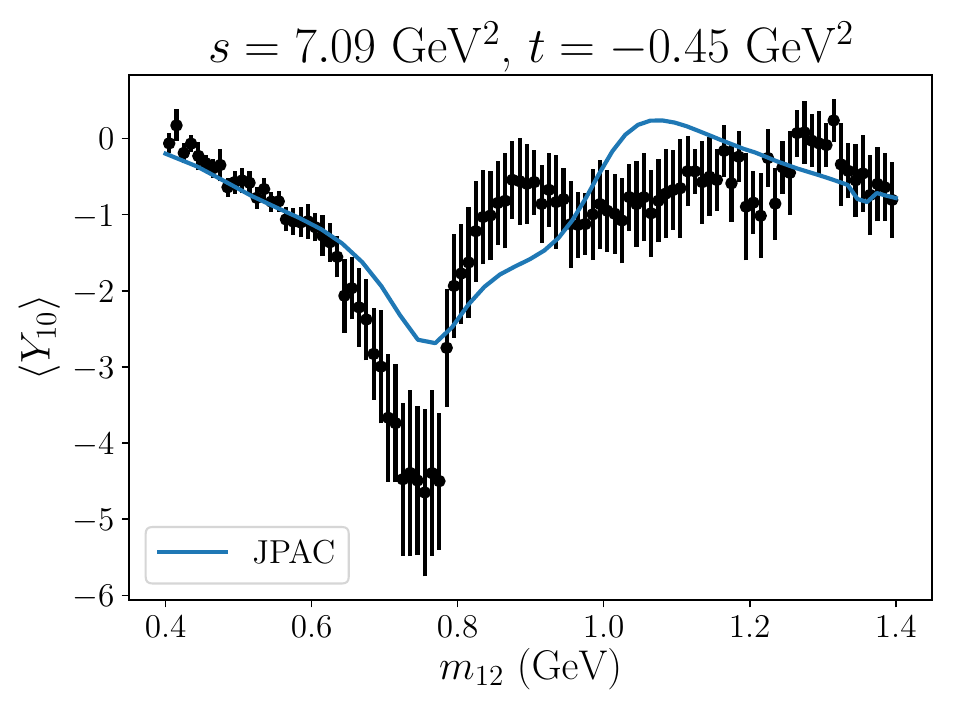}
\includegraphics[scale=0.25]{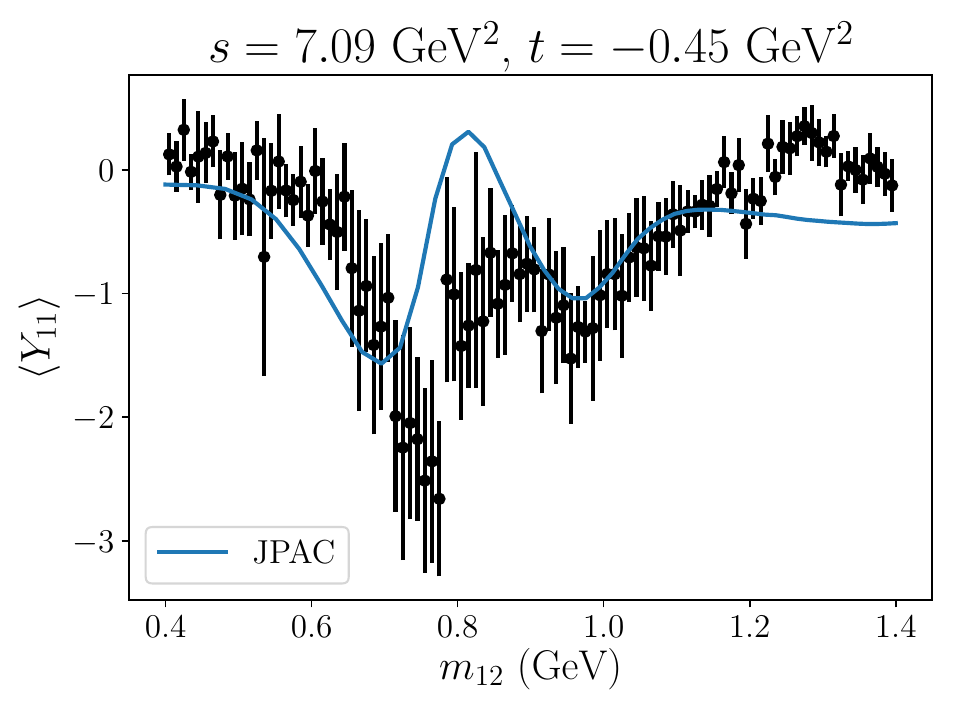}

\includegraphics[scale=0.25]{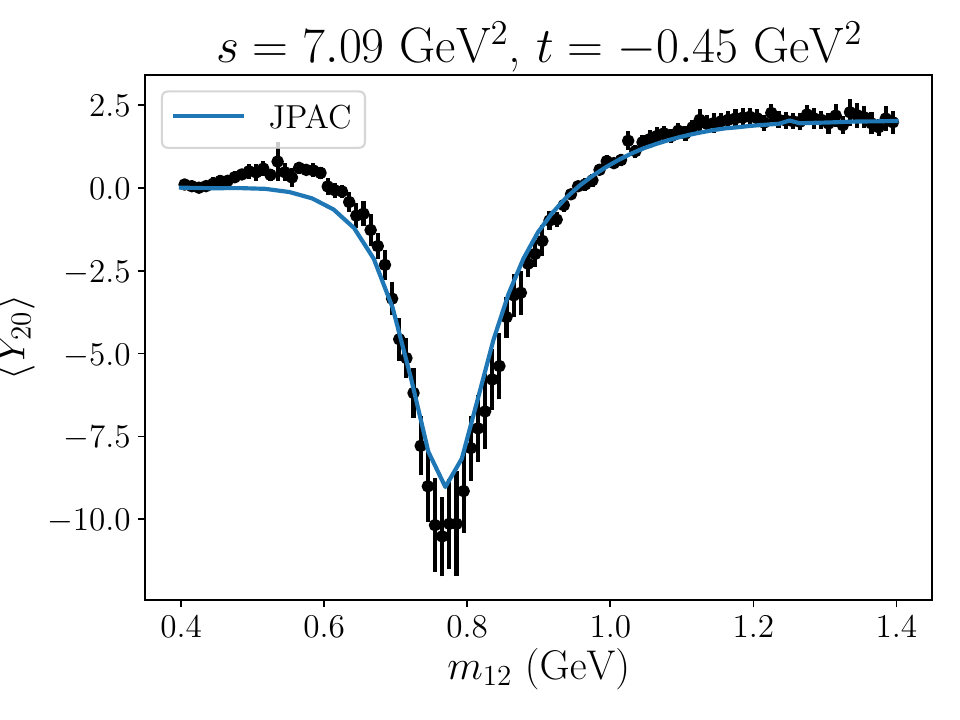}
\includegraphics[scale=0.25]{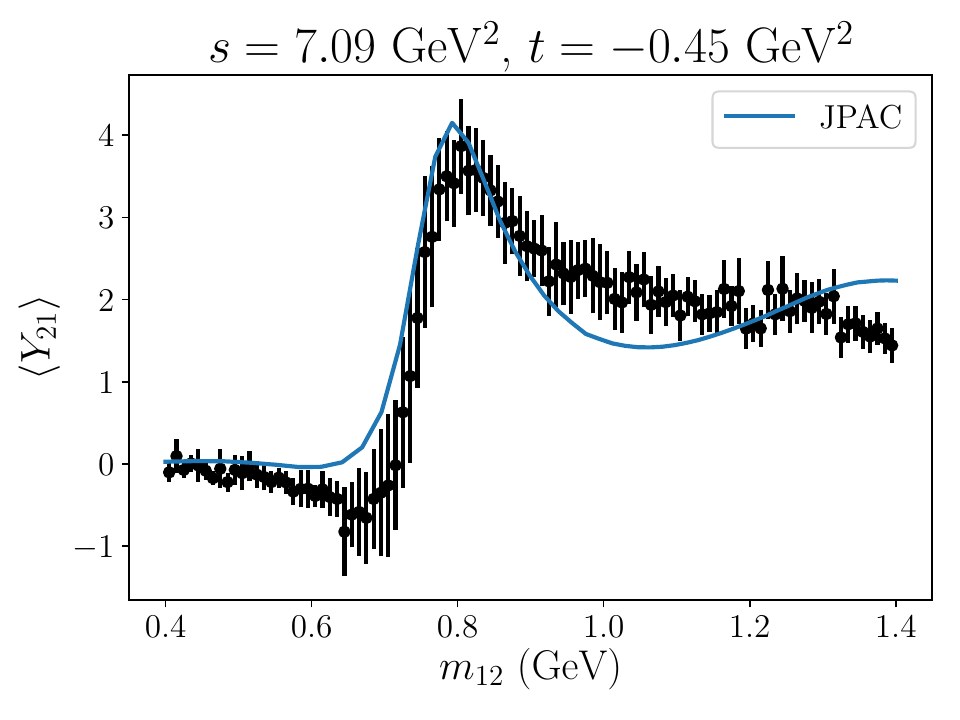}
\includegraphics[scale=0.25]{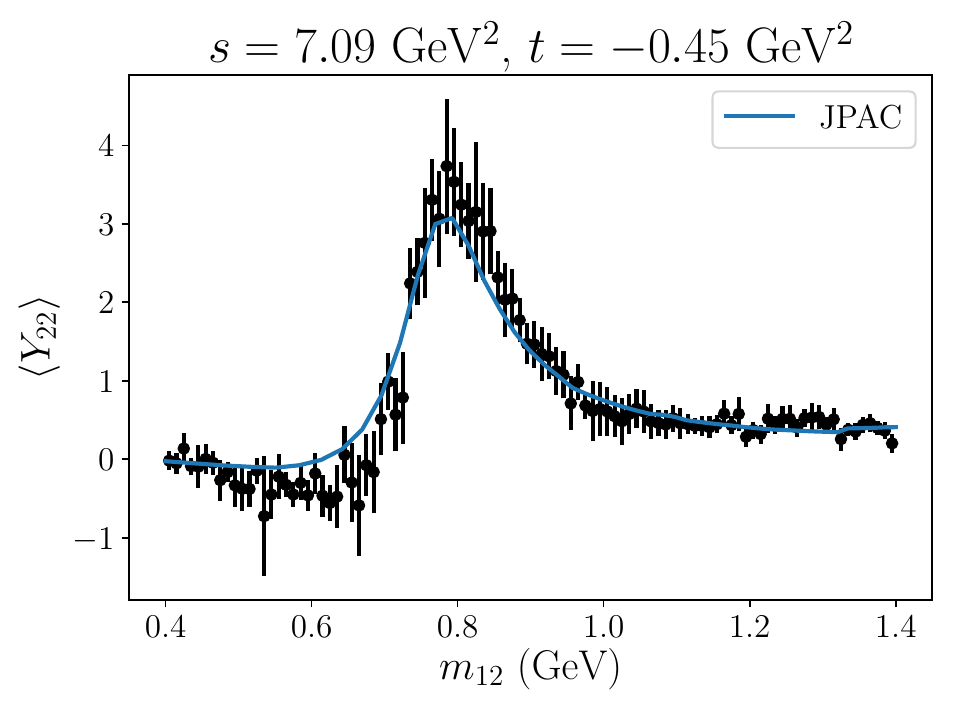}
\caption{Comparison of preliminary model predictions for angular moments with experimental results from CLAS~\cite{CLAS:2009ngd} at $E_\gamma=3.3~\si{GeV}$ and $t=-0.45~\si{GeV^2}$. }
\label{fig:predictions}
\end{figure}

\section{Results and Conclusion}
\label{sec:results}
A global fit of the six angular moments at is performed at fixed $s$ and $t$. The preliminary results of the model are shown in Fig.~\ref{fig:predictions}. A reasonable description of the lowest six angular moments is obtained. Fits of similar quality are obtained for the $-t\in \{0.55,0.65,0.75,0.85,0.95\}$. The model contains $2+2+2+6+6=18$ free parameters corresponding to the relative weights of the two resonant $S$-wave contributions and one resonant $P$-wave contribution, plus non-resonant $S$- and $P$-wave contributions which are modelled with polynomials in $s_{12}$. These polynomial non-resonant contributions are required to obtain a good fit to a subset of moments. In particular, this non-resonant component is required to describe $\expval{Y_{20}}$. The general agreement of the model with data gives confidence that the model captures the essential features of two-pion photoproduction in this kinematic region. Possible further research includes studying the $t$-dependence of the $g_{LM}$s and extrapolating the model to higher photon energies in order to make predictions for CLAS12 and GlueX.

This proceedings presented a theoretical model of two-pion photoproduction which incorporated the prominent $\rho(770)$ resonance and leading background contribution from the Deck mechanism. A good agreement between model and data was obtained after the relative weights of partial waves were fit to experimental data, suggesting that the model correctly identifies the production mechanisms and partial waves which are instrumental for the description of the angular moments.
This model can be applied to make predictions at photon energies relevant for the CLAS12 and GlueX experiments. 

\section*{Acknowledgements}
This work contributes to the aims of the USDOE ExoHad Topical Collaboration, contract DE-SC0023598. In addition, the support of project PID2020-118758GB-I00, financed by the Spanish MCIN/ AEI/10.13039/501100011033/ is acknowledged. The support of project No. 2018/29/B/ST2/
02576 (National Science Center), partly financed by a Polish research project is also acknowledged.

\end{fmffile}
\bibliography{bibliography}
\end{document}